\documentclass[12pt,preprint]{aastex}

\def\EQ{\begin{equation}}
\def\EN{\end{equation}}
\def\EQA{\begin{eqnarray}}
\def\ENA{\end{eqnarray}}
\usepackage{amsmath}
\usepackage{multirow}

\begin{document}

\title{On a stochastic model for the spin-down of solar type stars}
\author{\large Nicolas Leprovost and Eun-jin Kim}
\affil{Department of Applied Mathematics, University of Sheffield, Sheffield S3 7RH, UK}

\begin{abstract}
Modeling the rotation history of solar-type stars is still an unsolved problem in modern astrophysics. One of the main challenges is to explain the dispersion in the distribution of stellar rotation rate for young stars. Previous works have advocated dynamo saturation or magnetic field localization to explain the presence of fast rotators and star-disk coupling in pre-main sequence to account for the existence of slow rotators. Here, we present a new model that can account for the presence of both types of rotators by incorporating fluctuations in the solar wind. This renders the spin-down problem probabilistic in nature, some stars experiencing more braking on average than others. We show that random fluctuations in the loss of angular momentum enhance the population of both fast and slow rotators compared to the deterministic case. Furthermore, the distribution of rotational speed is severely skewed towards large values in agreement with observations.
\end{abstract}

\keywords{stars: evolution -- stars: interiors -- stars: rotation}

\maketitle

\section{Introduction}
To first approximation, the Sun at present is rotating as a solid body, the average rotation of the core and the envelope being roughly the same. However, this has not been always the case. During the Pre-Main Sequence (PMS) phase of their evolution, stars experience a contraction accompanied by changes in their internal structure. As a consequence, they spin-up and develop a radiative core which rotates faster than the envelope. If the coupling between the core and the envelope is sufficiently strong, the core can in turn accelerate the envelope. Towards the end of the PMS, the loss of angular momentum through stellar wind delays the spin-up of the convective envelope. In contrast, on the Main Sequence (MS), stars experience mostly a fast spin-down of the convective envelope, as the wind braking time-scale becomes the shortest scale \citep{Keppens95}. Note however that there is still small changes in the structure at the beginning of the MS.

Modeling the rotation history of solar-type stars is a formidable task due to a number of unknown physical processes to be understood. First, the loss of angular momentum through stellar wind requires a 3D model of the solar wind while only a limited amount of observational data (e.g. on geometry, time variation, etc.) are available. Secondly, one has to model the transfer of angular momentum between the convective zone of the star and its core which is not completely understood: it can be through transfer of mass but also through viscous-like transport (such as transport by waves or turbulence). The mass transport is dominant in the earlier evolution (during the PMS) due to the change in the stellar structure, whereas the viscous transport dominates on the MS. Once all these processes are properly modeled with appropriate initial conditions for the rotation rates of stars, it is possible to trace their rotation history to explain observations. The main observations to account for are those of equatorial rotational velocity ($v$) distributions (or rotational rate) of solar-type star clusters. For instance, $\alpha$Persei (age $\sim 50 \mbox{Myr}$) exhibits a tail of rapid rotators ($v \sin i > 50 \;  \mbox{km} \, \mbox{s}^{-1}$) accounting for 50\% of the stars and a small proportion (around 15\%) of slowly rotating rotators  ($v \sin i < 10 \; \mbox{km} \, \mbox{s}^{-1}$). At later time, in the Pleiades cluster (age $\sim 70 \mbox{Myr}$), observations show a predominance of slow rotators ($v \sin i > 20 \; \mbox{km} \, \mbox{s}^{-1}$) with a very small right tail of rapid rotators. At the Hyades ages ($\sim 600 \mbox{Myr}$), most stars are slow rotators ($v \sin i < 10 \; \mbox{km} \, \mbox{s}^{-1}$). Spin-down is also observed in times later than $600 \mbox{Myr}$, the rate of slow-down being given by the Skumanich relation \citep{Skumanich72}: $v \sin i \propto t^{-1/2}$ where $t$ is the age of the star.

Different mechanisms have been proposed to explain the spin-down of solar-type stars. As the loss of angular momentum is achieved through the magnetized stellar wind, an important ingredient is the prescription of the dependence of the magnetic field of a star on its rotation rate (the so-called dynamo relation). When this relation is linear (as suggested by stellar dynamo models), the decrease in rotation rate can however be shown to be too rapid in the early evolution and cannot thus account for the large tail of rapid rotators at the age of $\alpha$ Persei. Consequently, a number of models have invoked the saturation of the loss of angular momentum for rapidly rotating stars. In particular, \citet{Keppens95} and \citet{Barnes96} have assumed that the loss of angular momentum saturates for a rotation rate above a prescribed threshold, due to the saturation of dynamo, and obtained rotation distributions which agree reasonably well with observations for the value of threshold of $\Omega_t \sim 20 \Omega_\Sun$. The saturation of dynamo at $\Omega_t \sim 20 \Omega_\Sun$ can be supported by the observation of chromospheric activity (linked to the magnetic activity) which appears to flatten for a rotation rate of a similar order of magnitude. However, observations of star-spot coverage \citep{ODell95} seem to indicate a saturation for somewhat higher value of the rotation rate (typically $\Omega_t \sim 60-100 \Omega_\Sun$). It thus led other authors to suggest that the saturation of the angular momentum loss is due to a polar localization of the magnetic activity rather than the saturation of a dynamo process. Indeed, the localization of open magnetic field lines at higher latitude, which is observed for rapidly rotating stars, was shown to reduce the transport of angular momentum \citep{Buzasi97}, and thus to be an efficient mechanism for dynamo saturation. 

The purpose of this paper is to propose a new model of spin-down that can explain the existence of both fast and slow rotators in the early MS evolution. The key idea is based on a recent growing number of observational evidences for prominent intermittency in various solar/stellar activities, involving a broad range of temporal and spatial scales. For instance, in the case of the Sun, in addition to the 22-year activity cycle, variations on both shorter and longer time-scales than 22 years have been observed \citep{Forgacs07}. Similarly, the solar wind exhibits time-varying fluctuations which are far from negligible \citep{Bavassano05}. The presence of fluctuations in the stellar wind characteristics makes the loss of angular momentum a time variable process (intermittent). Consequently, being affected by fluctuations in the stellar wind, the loss of angular momentum cannot be properly modeled by using only the average rotation rate and magnetic field of the star. We thus treat the loss of angular momentum as a stochastic process whose statistical properties are to be investigated.

\section{Model}
To study stellar spin-down, we assume that both the core and the envelope rotate rigidly and that the coupling between the two is achieved through viscous-like transport mechanism \cite[for details, see][and references therein]{Keppens95}. The evolution of the angular momentum of the radiative core $J_c = I_c \Omega_c$  and envelopes $J_e = I_e \Omega_e$  can be written as:
\EQA
\label{System1}
\frac{dJ_c}{dt} &=& -  \dot{J}_{v} + \dot{J}_{m} \; , \\ \nonumber 
\frac{dJ_e}{dt} &=& \dot{J}_{v} - \dot{J}_{m} - \dot{J}_w  \; . 
\ENA
Here $J_c$, $J_e$ and $\Omega_c$, $\Omega_e$ are the angular momentum and angular rotation of the radiative core and envelope respectively. In the following, all our calculations are performed by using parameter values typical of the Sun with the moments of inertia: $I_c=59.87 \times 10^{52} \, \mbox{g} \mbox{cm}^2$ and $I_e=3.398 \times 10^{52} \, \mbox{g} \mbox{cm}^2$. $\dot{J}_m$ is the transfer of angular momentum associated with the mass exchange between the core and the envelope which is important primarily during the PMS phase. In the following, we focus on the evolution on the zero-Age Main Sequence (ZAMS) and consequently assume that $\dot{J}_m$ is negligible compared to the the exchange of angular momentum by visco-magnetic coupling mechanism between the core and the envelope $\dot{J}_{v}$. The latter is given by:
\EQ
\dot{J}_v = \frac{J_c I_c - J_e I_e}{\tau_c (I_c + I_e)} = \frac{I_c I_e}{\tau_c (I_c + I_e)} (\Omega_c - \Omega_e) \; ,
\EN
where $\tau_c$ is the coupling time between the radiative core and convection zone. In most previous studies, the coupling timescale $\tau_c$ was assumed to be fixed during the evolution of a star, for instance, with a value $\tau_c \sim 20 \, \mbox{Myr}$ \cite{Keppens95}. 

The braking due to the solar wind is assumed to be of the viscous type:
\EQ
\label{AMLoss}
\dot{J}_w = \frac{J_e}{\tau_w} \; .
\EN
To prescribe the braking time $\tau_w$, we use the Weber-Davis model \citep{Weber67} which has been shown to exhibit a transition between a slow magnetic rotator (SMR) and a fast magnetic rotator (FMR) regimes \citep{Belcher76}. In the FMR regime, the loss of angular momentum is mainly due to magnetic braking, whereas in the SMR regime, the main source of acceleration of the stellar wind is through thermal processes. Using asymptotic expression for the braking term given by \citet{MacGregor91}, the loss of angular momentum can be written as:
\EQ
\label{BrakingTime}
\frac{1}{\tau_w} = \frac{1}{\tau_{w\Sun}} \times
\begin{cases}
\left(\Omega_* B_* / \Omega_e \right)^{2/3} B^{4/3} & \mathrm{if} \quad \Omega_e > \Omega_* \\
B^2 & \mathrm{otherwise}
\end{cases}
\; .
\EN
Here $t_{w\Sun} = 300 \mbox{Myr}$ is the spin-down time of the present Sun; $\Omega_*$  (corresponding to a magnetic field $B_*$) is the threshold at which the transition between the SMR and the FMR regimes occurs. Eq. (\ref{BrakingTime}) is chosen to ensure continuity of $\tau_w$ at the transition point and to match the present spin-down time of the solar rotation. In all the calculations presented in the paper, we use the threshold value  $\Omega_*= 3.5 \Omega_\Sun = 1.05 \times 10^{-5} \, \mbox{s}^{-1}$. 

\begin{figure}[h]
\includegraphics[scale=0.5,clip]{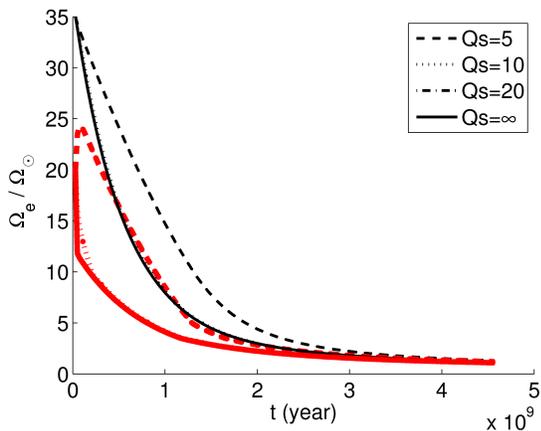}
\caption{\label{UniqueQs} Time-history of the angular rotation of the core (in thin lines) and the envelope (in thick lines) of a solar-type star for different values of the saturation threshold. $\Omega_\Sun$ is the present solar rotation rate.}
\end{figure}

As the loss of angular momentum is achieved through the magnetized stellar wind, the spin-down timescale $\tau_w$ depends on the rotation rate of the star and the magnetic field. To express $\tau_w$ in terms of rotation rate, one requires the dependence of the magnetic field on the rotation rate. This is the so-called dynamo prescription. To obtain this, most previous models assumed that the magnetic field depends only on the angular rotation of the convection zone, with the magnetic field varying linearly with the rotation rate as $B = B_\Sun \Omega_e / \Omega_\Sun$. This linear relation was however shown to lead to too rapid spin-down in early MS (when the rotation rate of the stars are high) and cannot explain the heavy tail of fast rotators at the age of $\alpha$ Persei as noted in the introduction. Consequently, other authors assumed that the dynamo (and thus the angular momentum loss) saturates at high rotation rate. For example, \citet{Keppens95} and \cite{Barnes96} assumed the following dependence of the magnetic field on the rotation rate:
\EQ
\label{SatDynamo}
B = B_\Sun \times
\begin{cases}
\Omega_e / \Omega_\Sun & \mathrm{if} \quad \Omega_e < Q_s \Omega_\Sun \\
Q_s & \mathrm{otherwise}
\end{cases}
\; .
\EN
To illustrate the effect of saturation, we computed the evolution of a solar-type star for different values of the saturation threshold $Q_s$ starting from ZAMS by using initial condition: $\Omega_e = 20 \Omega_\Sun$ and $\Omega_c = 35 \Omega_\Sun$. The rotation rates of the core and the envelope are shown in Figure \ref{UniqueQs}.
It can be seen that as the saturation threshold $Q_s$ is lowered, the spin-down of the envelope is delayed because the loss of angular momentum is reduced. For the lowest threshold ($Qs=5$) that is considered, we even observe a spin-up at early stage of the evolution as the loss of angular momentum is smaller than the acceleration provided by the fast rotating core. 

\section{Effect of stochastic fluctuations}
\label{StocEffect}
To investigate the effect of fluctuations in the stellar wind, we consider a time dependent loss of angular momentum by replacing Eq. (\ref{AMLoss}) by the following:
\EQ
\label{StochLoss}
\dot{J}_w = \frac{J_e}{\tau_w} \xi(t) \; , 
\EN
where $\tau_w$ is given by the Weber-Davis model (\ref{BrakingTime}) and $\xi(t)$ is a stochastic noise. The statistics of this noise is chosen to be a $\Gamma$ distribution so that fluctuations in solar wind velocity have  Gaussian distribution (see appendix \ref{NoiseStat} for details). Specifically, the distribution of the noise is taken to be:
\EQ
\label{DistribGamma}
P(\xi)=\frac{1}{N} \xi^{1/2} \exp\left[-\frac{3 \xi}{2}\right] \; .
\EN
Here, $N$ is a normalization factor. Note that the noise $\xi$ is defined only for positive values to ensure that the loss of angular momentum given by Eq. (\ref{StochLoss}) is always positive (i.e. momentum is extracted from the star by the solar wind). Furthermore, the distribution (\ref{DistribGamma}) is chosen such that the average loss of angular momentum is the same as the deterministic case since $\langle \xi \rangle =1$. Here the angular brackets denote the average over the statistics of the noise (see Appendix \ref{NoiseStat}).
  
\begin{figure}[h]
\includegraphics[scale=0.4,clip]{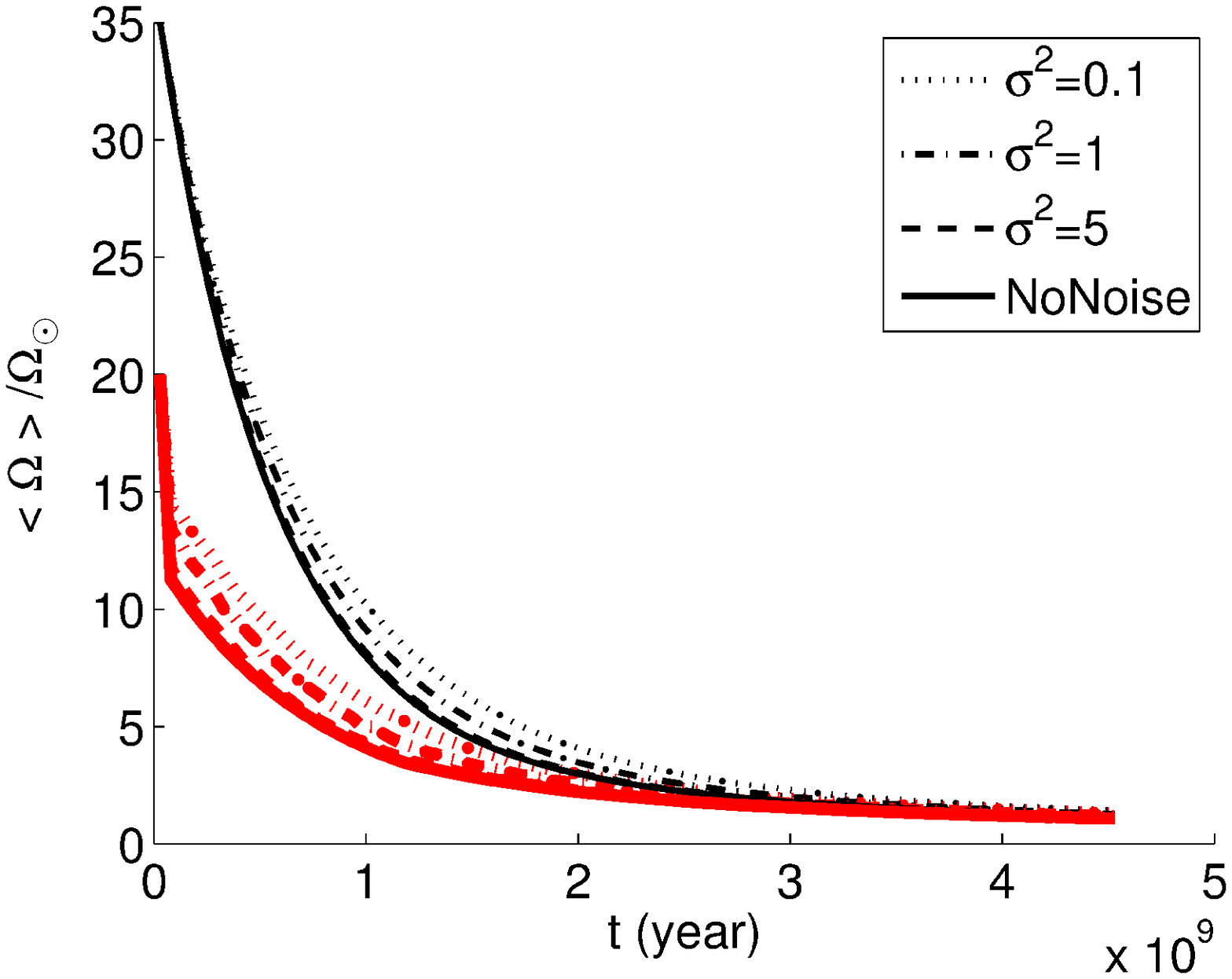}
\includegraphics[scale=0.4,clip]{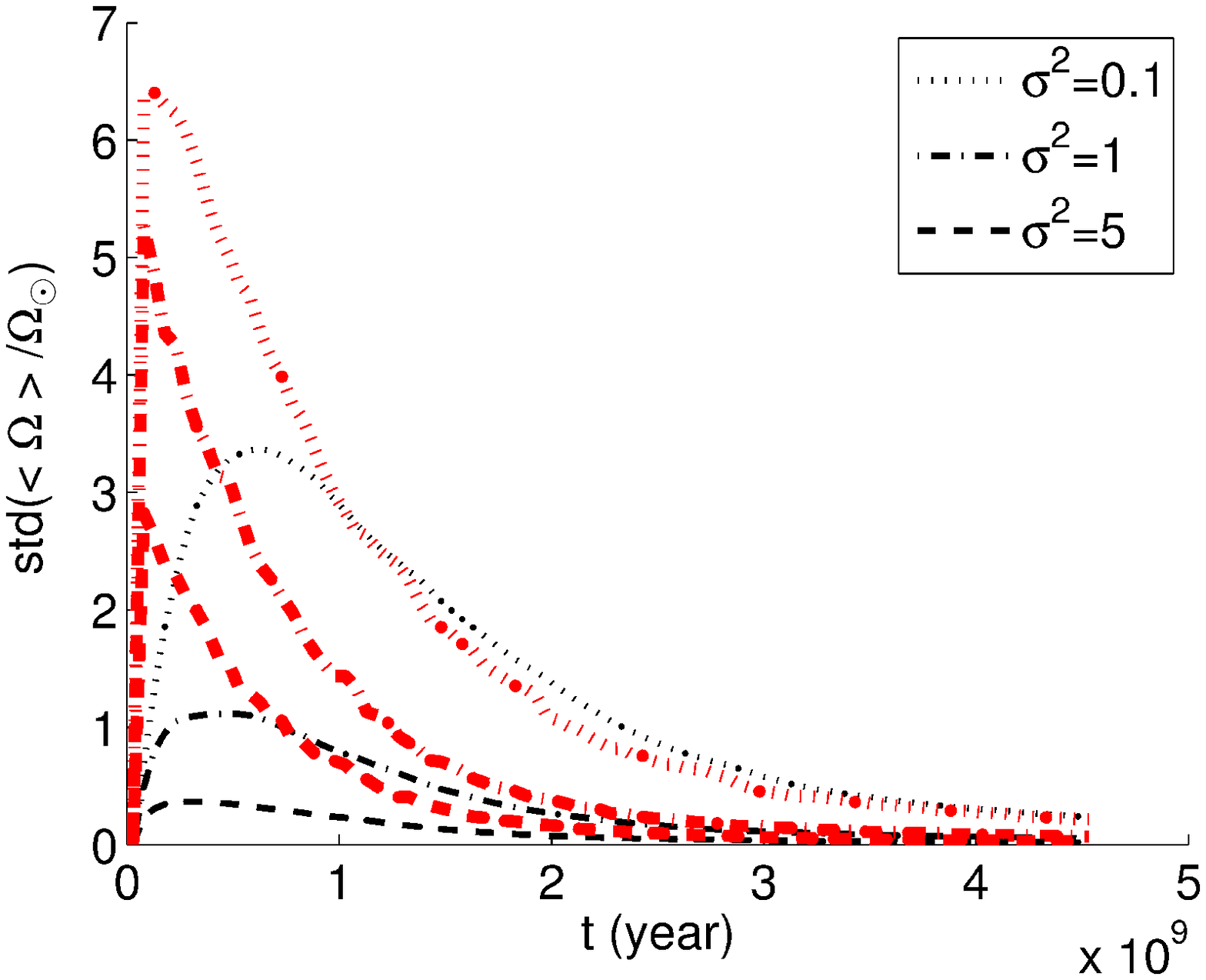}
\caption{\label{UniqueNoisyMean} Mean rotation rate of the core (in black) and the envelope (red) of a solar-type star for different values of the parameter $\sigma$. The initial conditions are $\Omega_e = 20 \Omega_\Sun$ and $\Omega_c = 35 \Omega_\Sun$.}
\end{figure}

\begin{figure}[h]
\includegraphics[scale=0.4,clip]{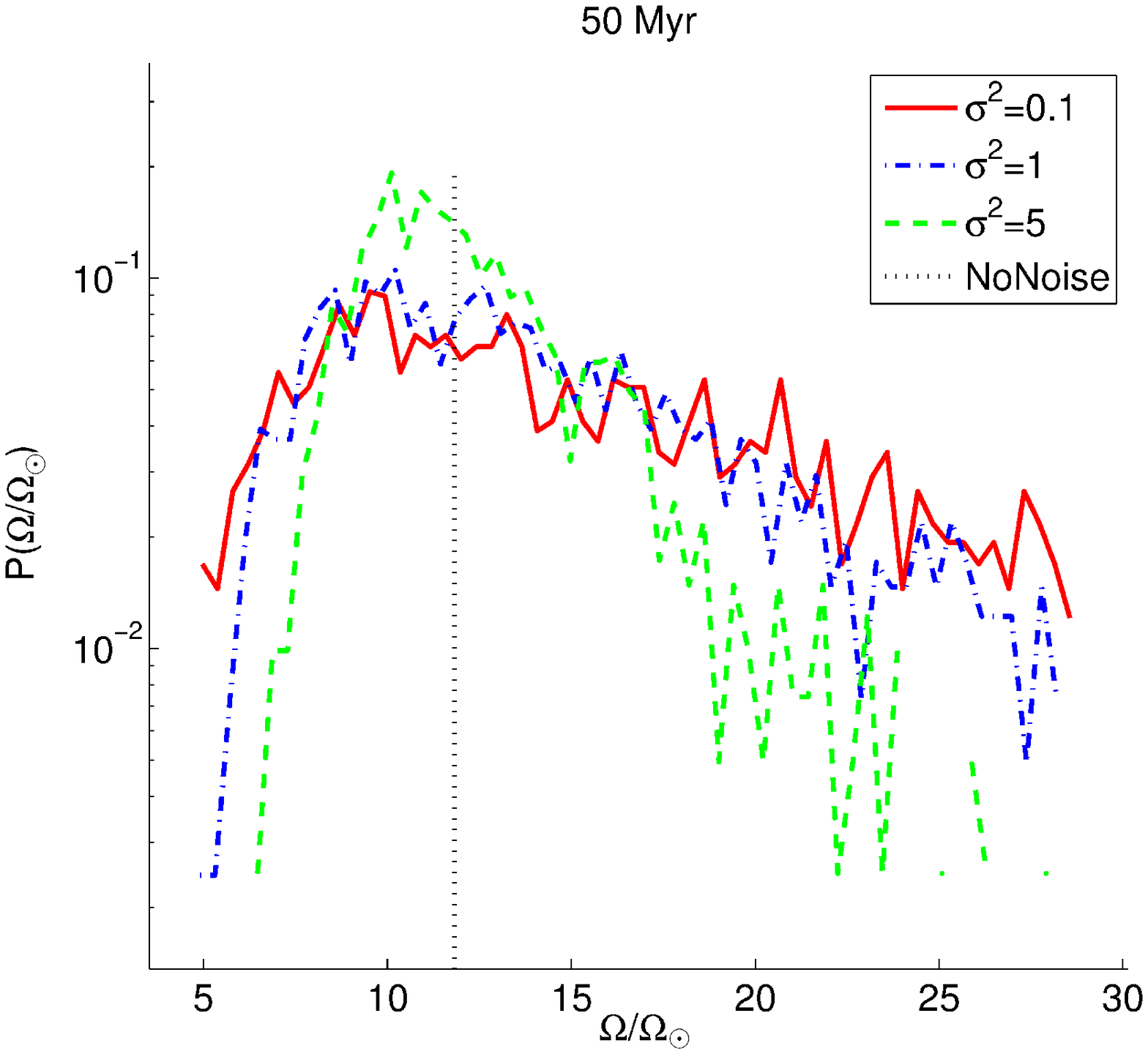}
\includegraphics[scale=0.4,clip]{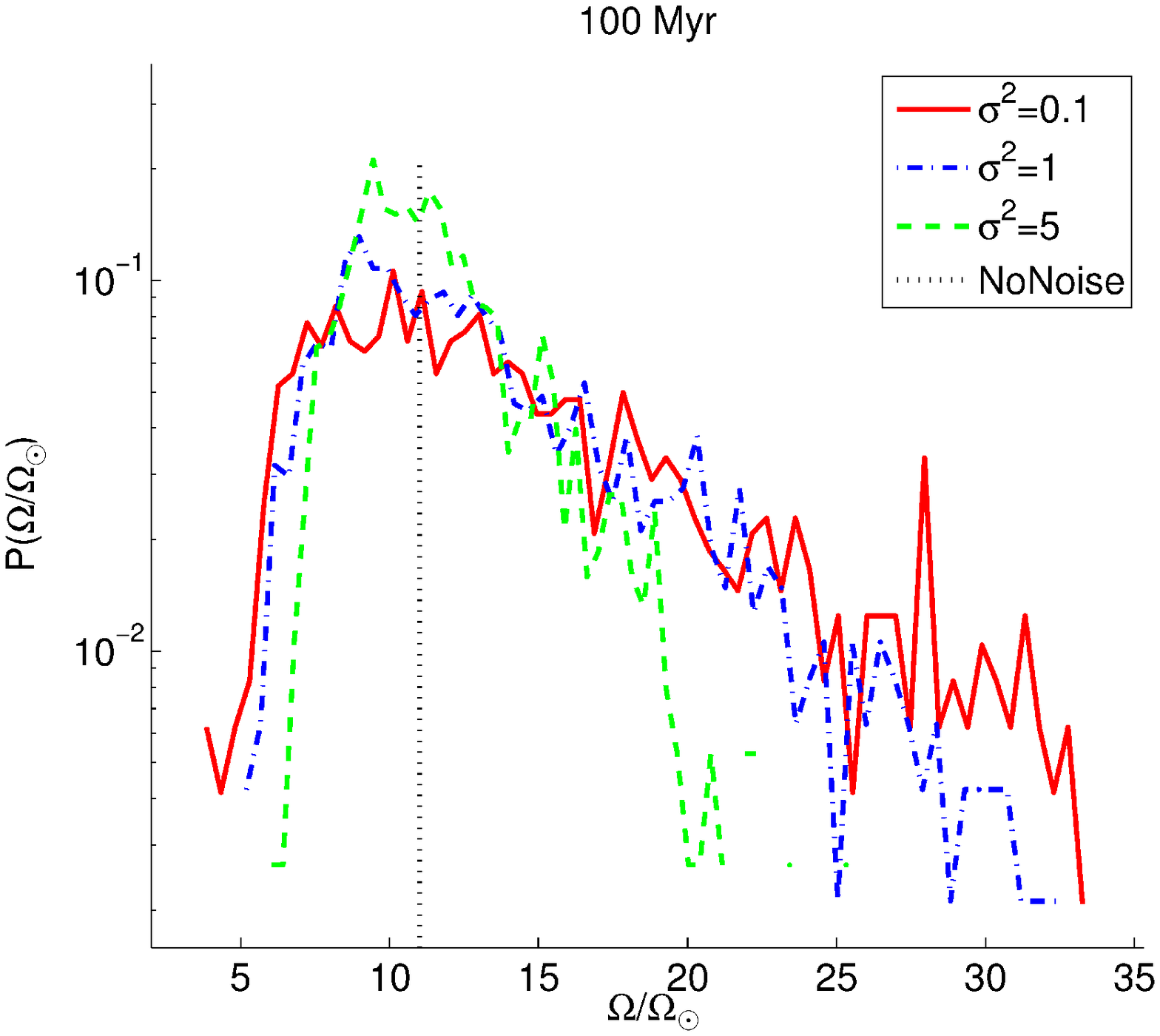}
\includegraphics[scale=0.4,clip]{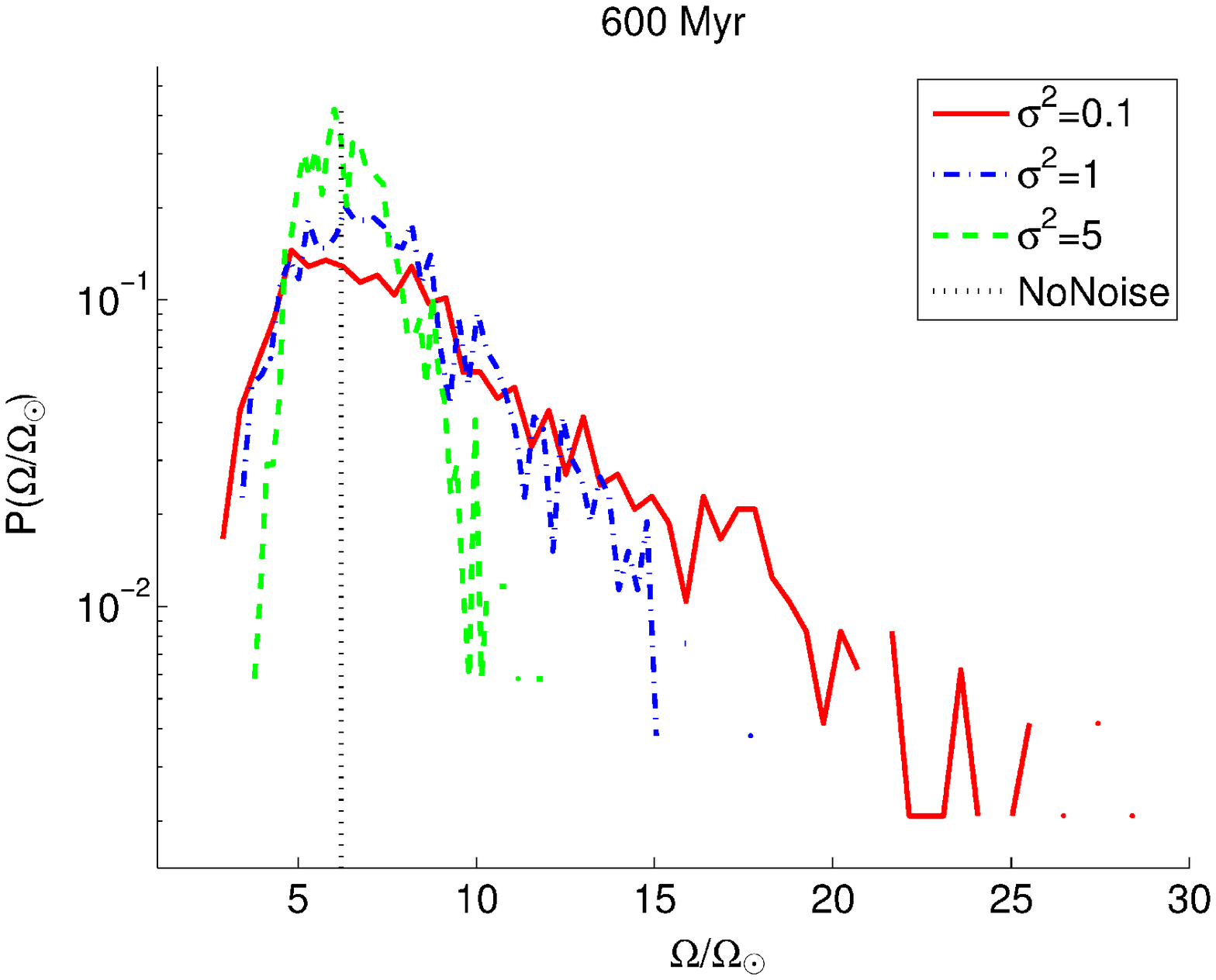}
\caption{\label{UniqueNoisyMyr} Probability distribution of the angular rotation of the envelope at different times. The initial condition is $\Omega_e = 20 \Omega_\Sun$ and $\Omega_c = 35 \Omega_\Sun$.}
\end{figure}

As the model is probabilistic, it is not possible to predict with certainty the rotation rate of the stars at any later times. We thus first examine the evolution of average rotation rate and standard deviation, which are shown in Figure \ref{UniqueNoisyMean}. It is seen that as the parameter $\sigma$ which depends on the correlation time of the noise (see Appendix \ref{NoiseStat}) is decreased, the spin-down of both the core and the envelope spin-down is reduced on average. 
Note that the mean rotation rate of the stars gives us only one measure of the rotational evolution on average. Thus, we compute the probability distribution function (PDF) of having a certain value of the angular rotation velocity of the envelope. Figure \ref{UniqueNoisyMyr} shows this PDF at three different times $t=50 \, , \, 100 \; \mbox{and} \; 600 \mbox{Myr}$. The main notable feature is that there is a large dispersion in the distribution around the mean value. Another important point is that the right tail is more pronounced than the left tail. This is due to the fact that the noise is multiplicative. That is, in Eq. (\ref{StochLoss}), the noise $\xi$  multiplies $J_e$ which is a function of $\Omega_e$. This makes the effect of the noise effectively stronger when $\Omega_e$ (and consequently $J_e$) is larger. 

\section{Rotation history of solar-type stars}
\label{CompKeppens}
In this section, we predict the distribution of angular velocity of solar-type stars observed in clusters by using our model. To investigate the effects of stochasticity in stellar winds on spin-down, we simulate the evolution on MS for a given distribution of initial rotation rates (for the core and the envelope) at ZAMS and by ignoring structural changes which are dominant mainly in PMS. To this end, we use the result of \citet{Keppens95} to fix our initial distribution at ZAMS=30Myr as follows. The distribution of envelope velocity is chosen to fit their results (see appendix \ref{InitKeppens} for details) while the ratio of core angular rotation to envelope rotation $\Omega_c / \Omega_e$ is fixed to be constant with the value $1.75$. After choosing the distribution (see appendix \ref{InitKeppens} for details), we take $5000$ initial conditions according to this distribution and evolve them by using both the deterministic and the stochastic models. In the following subsections, we compare our results with the observation of $\alpha$Persei (age $\sim 50 \mbox{Myr}$), the Pleiades cluster ($\sim 70 \mbox{Myr}$), and the Hyades Cluster ($\sim 600 \mbox{Myr}$).

\subsection{Evolution of the probability density with time}
\begin{figure}[h]
\includegraphics[scale=0.4,clip]{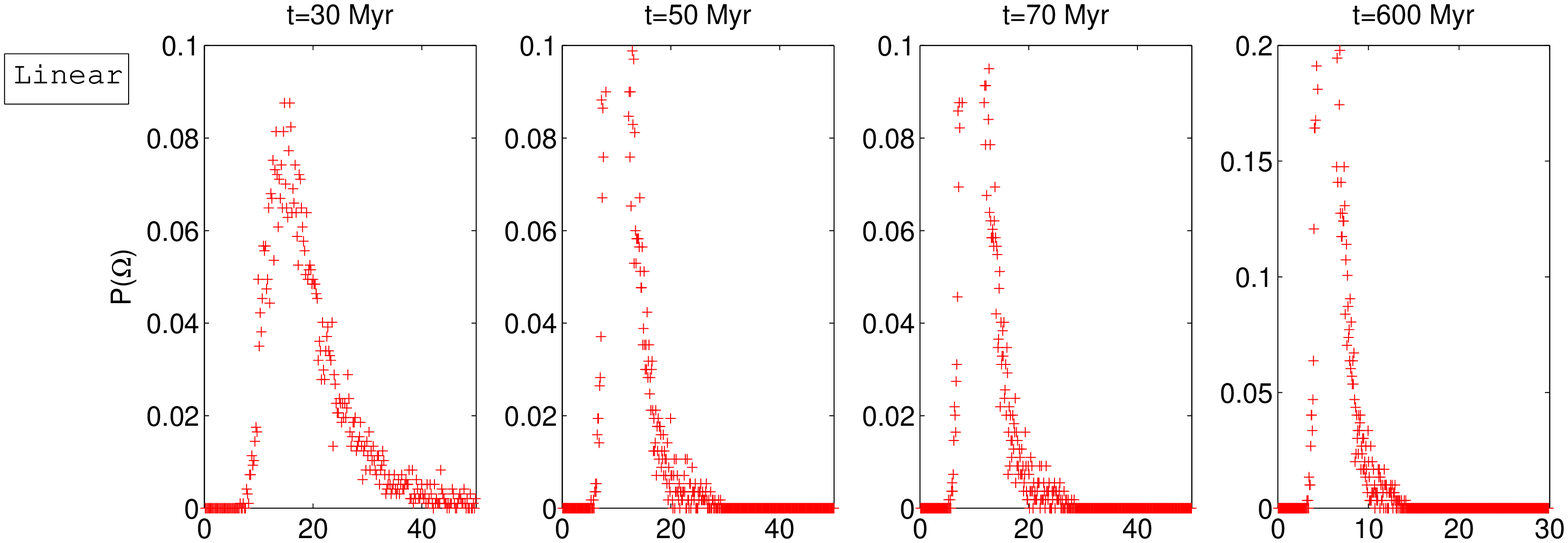} 
\includegraphics[scale=0.4,clip]{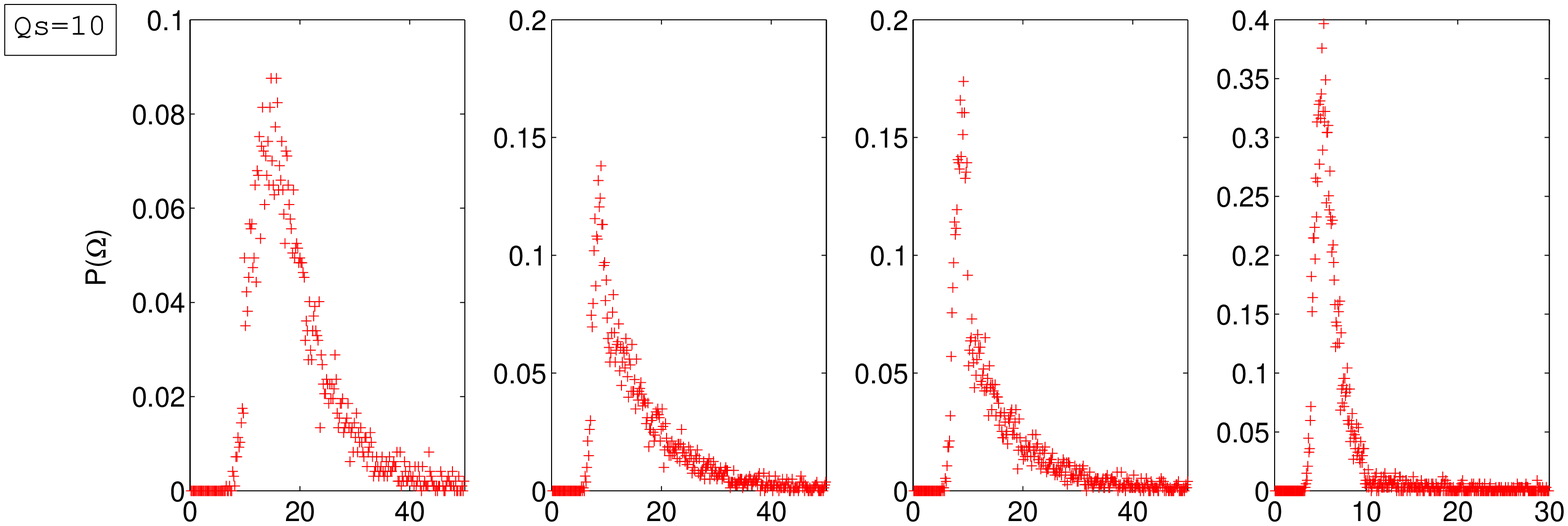}
\includegraphics[scale=0.4,clip]{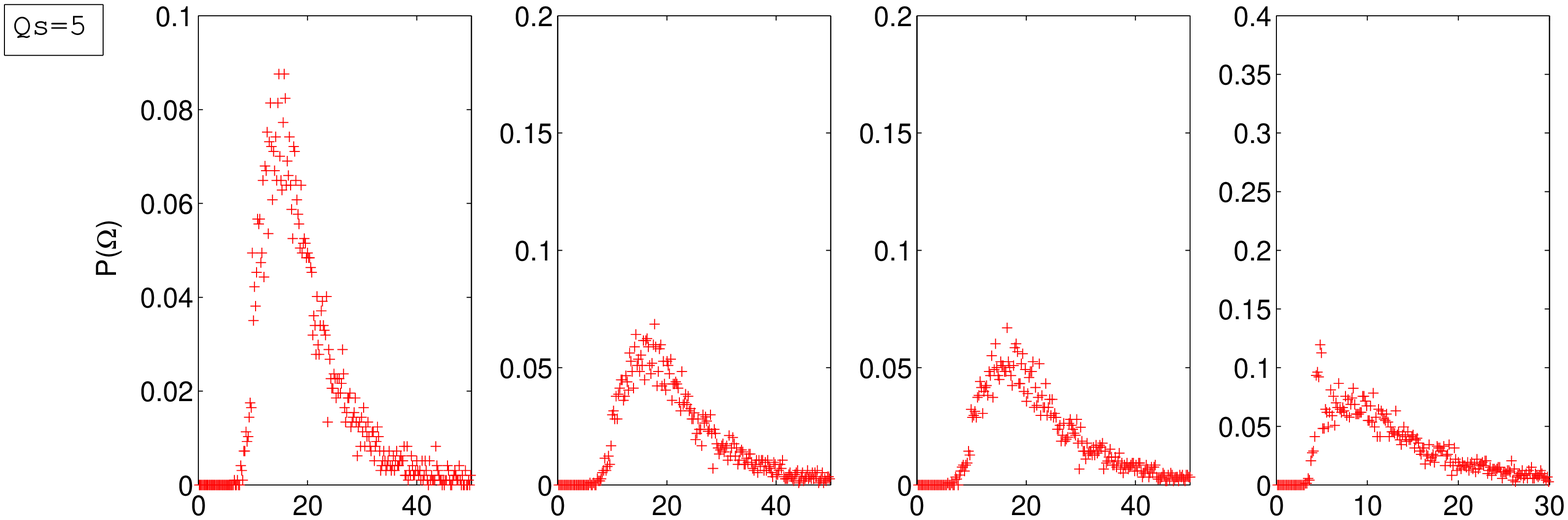}
\includegraphics[scale=0.4,clip]{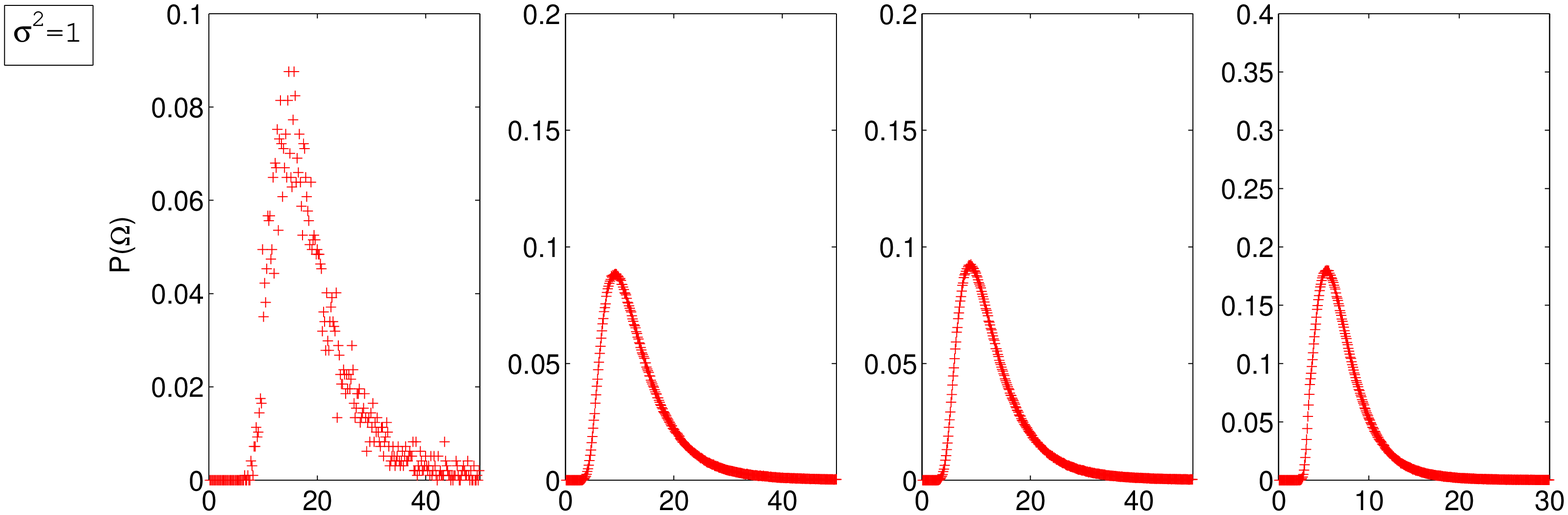}
\includegraphics[scale=0.4,clip]{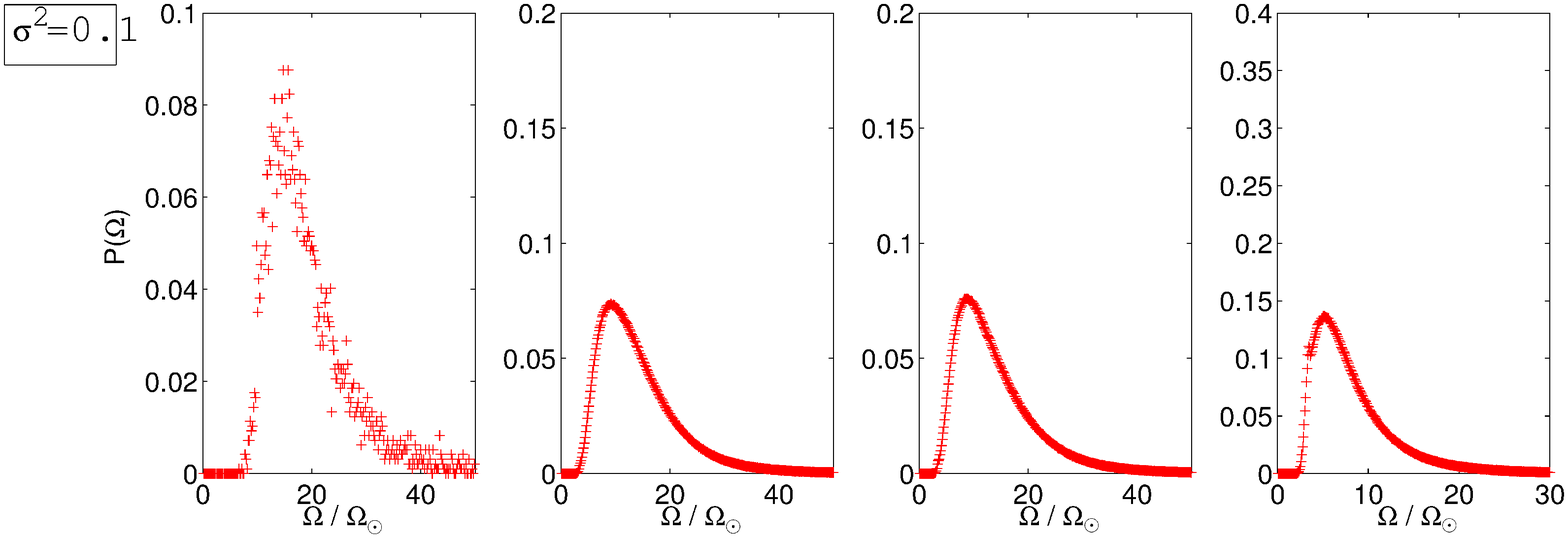}
\caption{\label{FigPDF} Evolution of the probability density with time for different values of the noise intensity and the saturation threshold. The top panel is for the case without saturation or noise, the second and third for two values of the saturation threshold and the last two for two values of the noise parameters. Recall that $\sigma$ is inversely proportional to the correlation time of the noise.}
\end{figure}

Figure \ref{FigPDF} shows the evolution of the distribution of star rotation rate in the linear case (without saturation or noise), the saturated case (with two different values of saturation threshold $Q_s=10$ and $Q_s=5$) and the stochastic case ($\sigma^2=1$ and $\sigma^2=0.1$). The top row shows that in the case without saturation nor noise, the entire distribution simply shifts to the left (smaller rotation rate) without dramatic modification in the shape of the distribution. In contrast, the next two rows show that, in the saturated case, the distribution develops a large right tail for rapid rotators. It is because these stars which rotate faster than the rotation threshold experience a weaker spin-down due to the saturation of magnetic field. The last two rows show the influence of the noise on the rotation distribution. In this case, the distribution also develops a right tail (high rotation) due to fluctuations in the spin-down. Recall that the tail consists of stars which have experienced less braking in their spin-down. The main difference between the saturated and stochastic cases is that the distribution in the stochastic case is a little shifted to the left compared to that in the saturated case, therefore accounting for a larger proportion of slow rotators. This is because a smaller percentage of stars experiences a weak braking in the stochastic case than in the saturated case. In contrast, the saturation mechanism affects solely the right tail as only rapidly rotating stars are prevented from spinning down. 

\clearpage

\subsection{Percentages of slow and fast rotators}
Table \ref{TableQs} shows the dependence on the saturation threshold and the noise intensity of the percentage of stars having an equatorial rotation $V_{eq} < 10 \; \mbox{km} \, \mbox{s}^{-1}$, $10 \; \mbox{km} \, \mbox{s}^{-1} < V_{eq} < 30 \; \mbox{km} \, \mbox{s}^{-1}$, $30 \; \mbox{km} \, \mbox{s}^{-1} < V_{eq} < 50 \; \mbox{km} \, \mbox{s}^{-1}$ and $V_{eq} > 50  \; \mbox{km} \, \mbox{s}^{-1}$ for ZAMS at three different times ($50 \, , \, 100 \; \mbox{and} \; 600 \, \mbox{Myr}$) corresponding to the age of the three clusters of interest. Note that the initial condition is fixed at $30 \mbox{Myr}$ to fit \citet{Keppens95} and is therefore the same in all simulations. Table \ref{TableQs} shows that the case $Q_s=20$ is very similar to the deterministic case without saturation nor noise. This is simply because the initial distribution is chosen such that most of the stars have rotation rate smaller than $20 \Omega_\Sun$. However, as the saturation threshold is lowered, the number of rapidly rotating stars is increased whereas the proportion of slow rotators is still $0 \%$ at $50 \mbox{Myr}$ and $70 \mbox{Myr}$. Only at later time $t=600 \mbox{Myr}$  is the effect of the saturation on the slow riotators visible. For $Q_s=5$, the number of slow rotators is significantly smaller than for the cases $Q_s=10$ and $Q_s=20$. 

Let us now examine the effect of noise for different values of the parameter $\sigma$ (which measures the correlation time of the noise). It can be seen from Table \ref{TableQs} that the case with the noise $\sigma^2=5$ is very similar to the case without saturation nor noise. When the parameter $\sigma$ is increased (corresponding to a decrease in the correlation time of the noise), the proportion of fast rotators is increasing. Furthermore, the  number of slow rotators is also increasing as the effect of the noise becomes more important (with a maximum proportion of $2 \%$ obtained for $\sigma^2=0.1$). It is also interesting to note that the proportion of stars at $600 \mbox{Myr}$ is roughly independent of the value of the noise parameter $\sigma$ (varying from $0.7\%$ to $10.1\%$ for $V_{eq} > 30$).

\begin{table}[h]
\begin{tabular}{|c|c|c|c|c|c|}
\hline
 &  & $t=30\, \mbox{Myr}$ &  $t=50\, \mbox{Myr}$  & $t=70\, \mbox{Myr}$ &  $t=600\, \mbox{Myr}$ \\ 
\hline
\hline
\multirow{4}*{Linear Case} & $V_{eq} < 10$ &  0 \%  &  0  \%  &   0 \%  &   20.3 \%   \\ \cline{2-6}
& $10 < V_{eq} < 30 $ &   29.5 \%  &  85.6 \%  &   88 \%  &  79.7 \%    \\ \cline{2-6}
& $30 < V_{eq} < 50 $ &   50.4 \%  &  13.6 \%  &  11.3 \%  &  0 \%    \\ \cline{2-6}
& $V_{eq} > 50 $ &  20.1 \%  &  0.8 \%  &  0.6 \%  & 0 \%   \\ 
\hline
\hline
\multirow{4}*{$Q_s = 20$} & $V_{eq} < 10$ &  0 \%  &  0  \%  &   0 \%  &   22.4 \%   \\ \cline{2-6}
& $10 < V_{eq} < 30 $ &   29.5 \%  &  85.6 \%  &   87.8 \%  &  77.5 \%    \\ \cline{2-6}
& $30 < V_{eq} < 50 $ &   50.4 \%  &  11.8  \%  & 10.5 \%  &  0.1 \%    \\ \cline{2-6}
& $V_{eq} > 50 $ &  20.1 \%  &  2.7 \%  &  1.7 \%  & 0 \%   \\ 
\hline
\hline
\multirow{4}*{$Q_s =10$} & $V_{eq} < 10$ &  0 \%  &  0  \%  &   0 \%  &   22.6 \%   \\ \cline{2-6}
& $10 < V_{eq} < 30 $ &   29.5 \%  &  58.3 \%  &   65.2 \%  &  73.7 \%    \\ \cline{2-6}
& $30 < V_{eq} < 50 $ &   50.4 \%  &  27 \%  &  22.1 \%  &  2 \%    \\ \cline{2-6}
& $V_{eq} > 50 $ &  20.1 \%  &  14.7 \%  &  12.7 \%  & 1.7 \%   \\ 
\hline
\hline
\multirow{4}*{$Q_s =5$} & $V_{eq} < 10$ &  0 \%  &  0 \%  &   0 \%  &   7.7 \%   \\ \cline{2-6}
& $10 < V_{eq} < 30 $ &   29.5 \%  &  22.5 \%  &   20.9 \%  &  57.1 \%    \\ \cline{2-6}
& $30 < V_{eq} < 50 $ &   50.4 \%  &  45.6 \%  &  44.1 \%  &  21.7 \%    \\ \cline{2-6}
& $V_{eq} > 50 $ &  20.1 \%  &  31.9 \%  &  35 \%  & 13.5 \%   \\ 
\hline
\hline
\multirow{4}*{$\sigma^2 =5$} & $V_{eq} < 10$ &  0 \%  &  0.1  \%  &  0.2 \%  &  26.1 \%   \\ \cline{2-6}
& $10 < V_{eq} < 30 $ &   29.5 \%  &   77.8 \%  &   80.8 \%  &  73.2 \%    \\ \cline{2-6}
& $30 < V_{eq} < 50 $ &   50.4 \%  &  19.2 \%  &   16.7 \%  &  0.7  \%    \\ \cline{2-6}
& $V_{eq} > 50 $ &  20.1 \%  &  2.9 \%  &   2.4 \%  &  0 \%   \\ 
\hline
\hline
\multirow{4}*{$\sigma^2 =1$} & $V_{eq} < 10$ &  0 \%  &  0.9  \%  &   1.2 \%  &   21.8 \%   \\ \cline{2-6}
& $10 < V_{eq} < 30 $ &   29.5 \%  &  65.8 \%  &   68 \%  &   74.6 \%    \\ \cline{2-6}
& $30 < V_{eq} < 50 $ &   50.4 \%  &  25.8 \%  & 24 \%  &  3.3  \%    \\ \cline{2-6}
& $V_{eq} > 50 $ &  20.1 \%  &  7.4 \%  &  6.8 \%  &  0.3 \%   \\ 
\hline
\hline
\multirow{4}*{$\sigma^2 =0.1$} & $V_{eq} < 10$ &  0 \%  &  2  \%  &   2.2 \%  &   22.4\%   \\ \cline{2-6}
& $10 < V_{eq} < 30 $ &   29.5 \%  &  59.2 \%  &   60.4 \%  &   67.5 \%    \\ \cline{2-6}
& $30 < V_{eq} < 50 $ &   50.4 \%  &  28.8 \%  &  27.2 \%  &   8.2 \%    \\ \cline{2-6}
& $V_{eq} > 50 $ &  20.1 \%  &  10 \%  &  10.2 \%  &  1.9 \%   \\ 
\hline
\end{tabular}
\caption{\label{TableQs} Percentages of stars having a certain angular velocity for different values of the saturation rate $Q_s$ and the noise parameter $\sigma$. The other parameters have been fixed to $\tau_w=300 \mathrm{Myr}$ and $\tau_c=20 \mathrm{Myr}$.}
\end{table}

\section{Conclusion}
In this paper, we presented a new model of spin-down that can account for the existence of both fast and slow rotators at the early stage of rotational evolution on the MS. By including the crucial effect of random fluctuations in the loss of angular momentum, we formulated the solar spin-down as a stochastic process. We then performed numerical simulations of this stochastic system and showed that the distribution of stars rotation rate has a wide dispersion severely skewed towards high rotation rates. Mathematically, this follows from the property of multiplicative noise used here since the rotation rate multiplies the stochastic fluctuations. In this case, the effect of noise is more important for large rotation values, therefore severely increasing the right tail of the distribution.

One of our key results is that this model can successfully reproduce a large proportion of fast rotators at the early stage of the stellar evolution. In that respect, it has a similar effect to the saturation of the dynamo process which prevents the spin-down of fastest rotating stars. Unlike saturation (and the linear model), the stochastic model can also explain the presence of slow rotators in the left tail. This can be seen from Table \ref{TableQs} where the percentage of slow rotators is not zero in contrast to the linear and saturated dynamo. However, the amount of slow rotators (at maximum 2\%) is still smaller than the amount implied by observations (around 15\%). However, it should be noted that these observations overestimate the number of slow rotators as they give a measure of the projected equatorial velocity $ v_{eq} \sin i$ which gives only lower bounds on the equatorial velocity of stars. Therefore, the number of slow rotators in stellar clusters is not as a severe constraint as the number of fast rotators. To quantify this effect, we investigated in appendix \ref{AngleEffect} how the proportion of fast and slow rotators would be changed if the angle $i$ was taken into account by assuming a given distribution of angle at which the stars are seen. 

We emphasize that the main purpose of this paper is to pinpoint a novel effect of stochasticity (which has recently been observed in temporal evolution of the solar wind) on the dispersion in rotation rate of solar-type stars. To obtain better agreement with observations, it will be of interest to be fine-tune our model by incorporating other mechanisms or adjusting parameters. Some possibilities include:
\begin{itemize}
\item Different initial distributions. For instance, it has been proposed that the distribution at ZAMS should be bimodal due to a number of stars being prevented from spin-up during the PMS due to coupling with their accretion disk.
\item A combined effect of stochasticity and saturation.
\item Different values of the ratio of the core to envelope rotation rates at ZAMS.
\item Modeling of the momentum transfer between the core and the envelope by accounting for non-linear dependence on the differential rotation. For instance, shear instability can occur when the differential rotation is large enough, by increasing the momentum transport at the interface.
\item Different magnetic field configurations (e.g. dipolar) and distributions.
\end{itemize} 
These issues will be addressed in future contributions.

\begin{acknowledgments}
We thank K.B. MacGregor and M.J. Thompson for useful comments. This work was supported by U.K. STFC Grant No. ST/F501796/1. 
\end{acknowledgments}

\begin{appendix}
\section{Stochastic model}
\label{NoiseStat}
According to the Weber-Davis model, the loss of angular momentum due to solar wind can be expressed as follows \citep[see][for instance]{Keppens95}:
\EQ
\dot{J}_w \propto \frac{r_A^2 \dot{M}}{I_e} J_e \; ,
\EN
where $\dot{M}$ is the mass flux from the star and $r_A$ is the Alfv\'en radius (radius at which the wind flow speed equals the Alfv\'en speed). It can also be shown that the Alfv\'en radius is proportional to the intensity of the flow. Then, assuming that the fluctuations in the velocity are Gaussian, the distribution of the fluctuations in the angular momentum can be shown to be distributed as:
\EQ
\label{GammaDistrib}
P(m \propto r_A^2) \propto m^{1/2} \exp[-c m] \; .
\EN 
To ensure that the average loss of angular momentum is the same as that in the deterministic case, we require the distribution (\ref{GammaDistrib}) to have a mean value of $1$, by fixing the value of $c$ to be $c=3/2$. 

\begin{figure}[h]
\includegraphics[scale=0.37,clip]{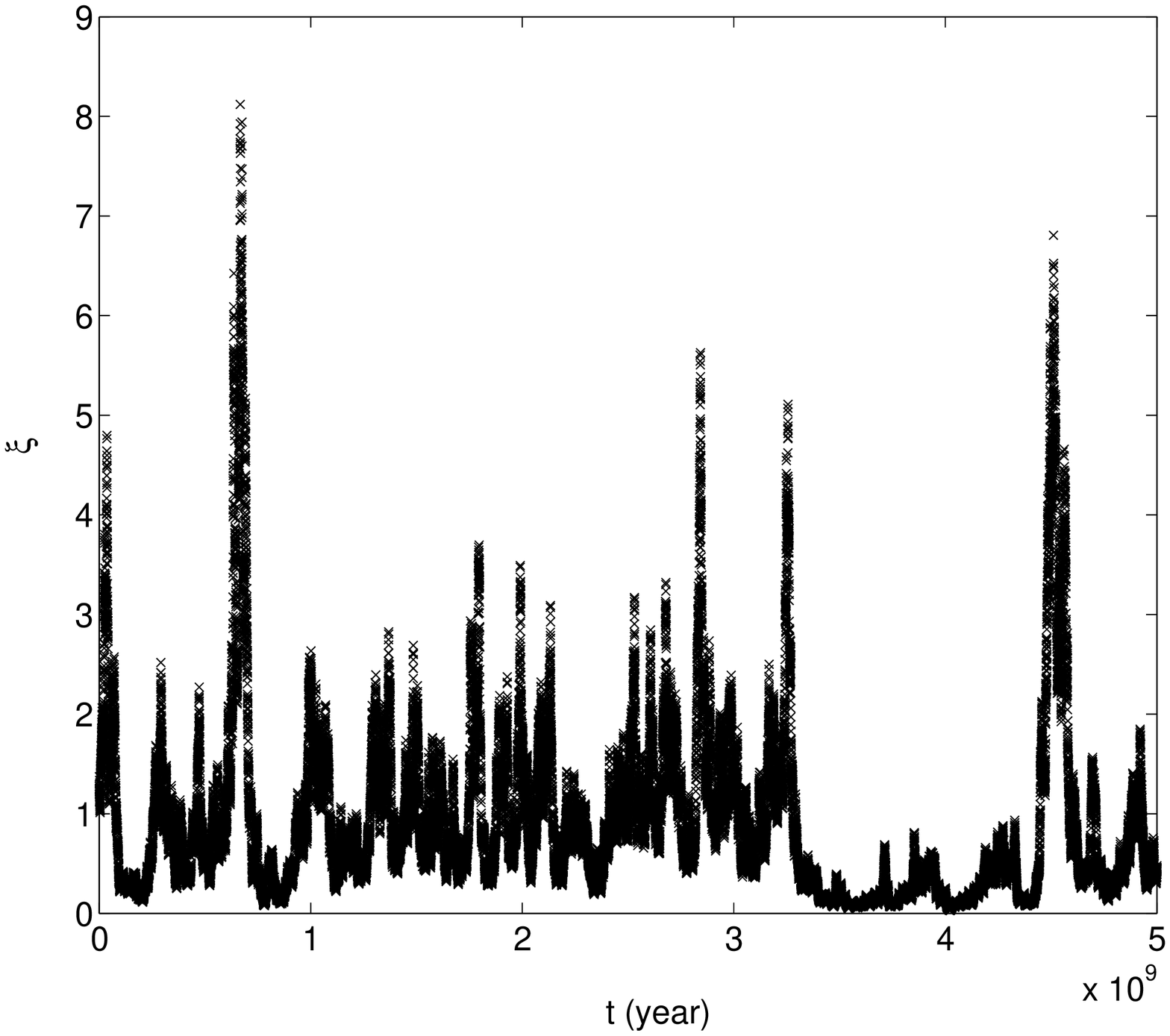}
\includegraphics[scale=0.37,clip]{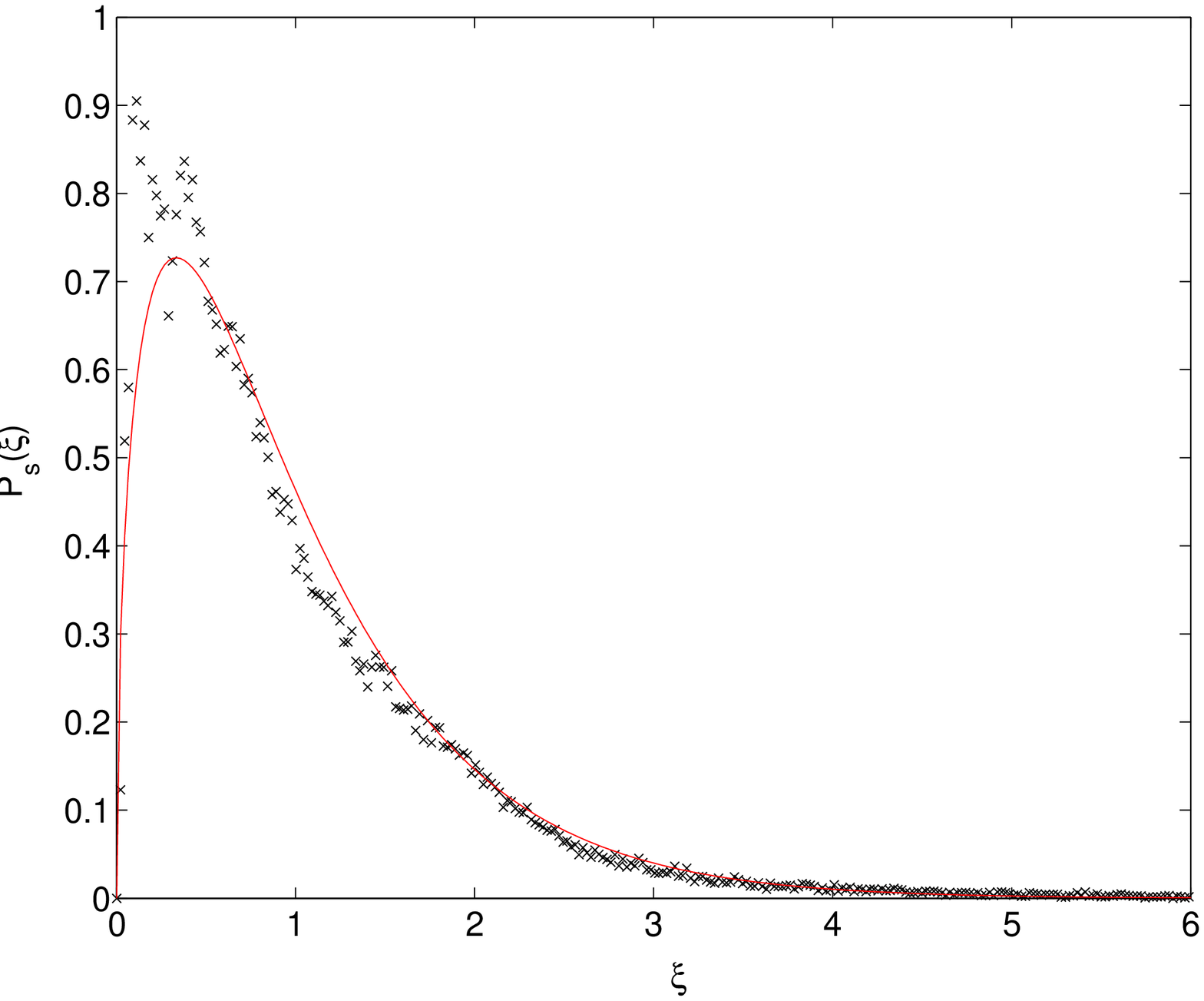}
\caption{\label{StocModelFig} Time series (left) of the noise used in the stochastic simulations and the corresponding probability distribution (right). The distribution (\ref{GammaDistrib}) is plotted as a solid line on the right panel.}
\end{figure}

\begin{figure}[h]
\includegraphics[scale=0.37,clip]{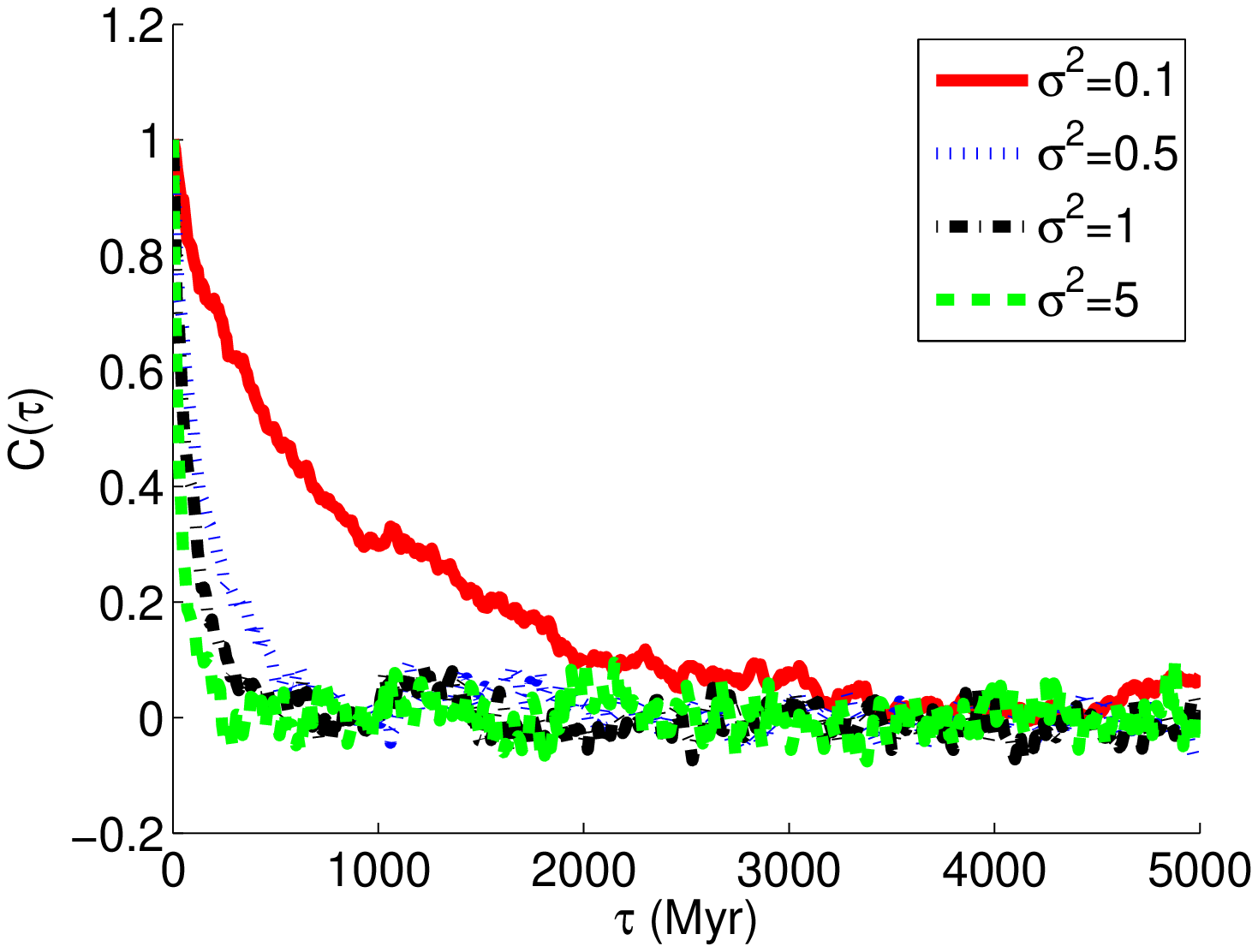}
\includegraphics[scale=0.37,clip]{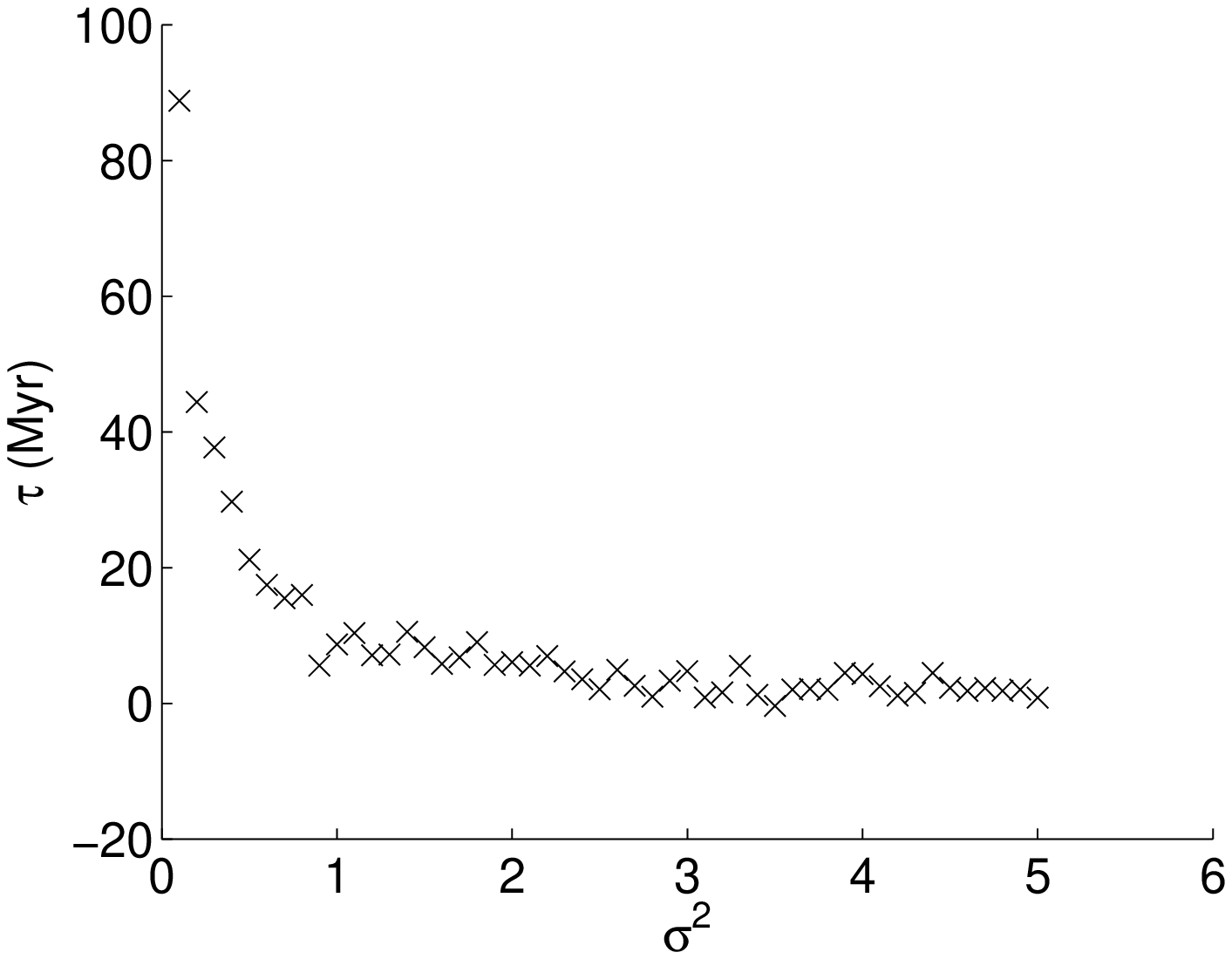}
\caption{\label{NoiseCorrelation} (Left) Temporal autocorrelation (\ref{CorrFunction}) of the noise for different values of the parameter $\sigma$. (Right) Correlation time of the noise as defined in Eq. (\ref{CorrTime}) as a function of $\sigma^2$.}
\end{figure}

To generate a time series of noise $\xi$ distributed according to Eq. (\ref{GammaDistrib}), we numerically solve the following stochastic differential equation:
\EQ
\label{StochasticModel}
\dot{\xi} = a \xi - g \xi^2 + \xi \Gamma(t) \; . 
\EN
Here $\Gamma$ is a Gaussian white noise with the correlation function: $\langle \Gamma(t) \Gamma(t') \rangle = 2 \sigma^2 \delta(t-t')$. We choose a random initial condition $\xi(t_0) > 0$; as $\xi=0$ is an absorbing point of Eq. (\ref{StochasticModel}), $\xi$ (and thus the loss of angular momentum) is always non-negative. Note that Eq. (\ref{StochasticModel}) is written assuming the Stratonovitch convention \citep{Kloeden92}. Using standard techniques, it is straightforward to show that the probability distribution $P(\xi,t)$ satisfies the following Fokker-Planck equation:
\EQ
\label{StochasticModel2}
\partial_t P = - \partial_\xi \left[\left(a \xi - g\xi^2\right)\right] + \sigma^2 \partial_\xi \left[\xi \partial_\xi \left(\xi P \right) \right] \; . 
\EN 
For sufficiently long time, the distribution $P$ in Eq. (\ref{StochasticModel2}) converges towards the stationary distribution $P_s$:
\EQ
\label{StatDistrib}
P_s(\xi) = \xi^{a/\sigma^2-1} \exp\left[- \frac{g \xi}{\sigma^2} \right] \; .
\EN
This stationary distribution  can be chosen to match Eq. (\ref{GammaDistrib}) with $c=3/2$ by taking $a / \sigma^2 = g / \sigma^2 = 3/2$. We have three parameters $(a,g,\sigma)$ and two relations which leave only one free parameter. We choose to vary $\sigma$ and fix the other two as $a = g = 3/2 \sigma^2 $. Eq. (\ref{StochasticModel2}) is then integrated until the stationary distribution is reached; results are shown in Figure \ref{StocModelFig}.
It is interesting to note that regardless of the value of $\sigma$, the noise in the stationary regime is distributed according to Eq. (\ref{GammaDistrib}). However, the autocorrelation function defined as:
\EQ
\label{CorrFunction}
C(\tau) = \langle (\xi(0) - \langle \xi \rangle)  (\xi(\tau) - \langle \xi \rangle) \rangle \; ,
\EN
is different for different values of the parameter $\sigma$ as shown on the left panel of Figure \ref{NoiseCorrelation}. The right hand panel shows the correlation time computed as:
\EQ
\label{CorrTime}
\tau_c = \int_0^{\infty} \frac{C(\tau)}{C(0)} d \tau \; ,
\EN
which is a decreasing function of $\sigma$ as shown on the left panel of Figure \ref{NoiseCorrelation}: the larger $\sigma$, the shorter correlated the noise. For instance, the correlation time is $89$, $21$, $8$ and $0.8 \mbox{Myr}$   for $\sigma^2=0.1,0.5,1$ and $5$, respectively. 

To simplify the formulation of the stochastic model, we use the following non-dimensionalised variables:
\EQA
x_1 = \frac{J_c}{(I_c+I_e) \Omega_\Sun} \; , \\ \nonumber
x_2 = \frac{J_e}{(I_c+I_e) \Omega_\Sun} \; , 
\ENA
where $\Omega_\mathrm{\Sun} = 3\times 10^{-6} \, \mbox{s}^{-1} $ is the present solar angular rotation. The two shell model (\ref{System1}) coupled with the stochastic model (\ref{StochasticModel}) can then be rewritten as:
\EQA
\label{System2}
\frac{d x_1}{dt} &=& - \alpha \,  x_1 + \beta \, x_2 \; ,\\ \nonumber 
\frac{d x_2}{dt} &=& \alpha \, x_1 - \beta \, x_2 - \gamma(x_2,B) x_2 \xi(t) \; , \\ \nonumber
\frac{d \xi} {dt} &=& \frac{3 \sigma^2}{2} (\xi - \xi^2) + \xi \, \Gamma(t) \; . 
\ENA
Here:
\EQ
\alpha_0 = \frac{I_e}{I_c+I_e} \; , \qquad \alpha = \frac{\alpha_0}{\tau_c} \; , \qquad \beta= \frac{1-\alpha_0}{\tau_c}  \quad  \mathrm{and}
\quad \gamma(x_2,B) = \frac{1}{\tau_w(x_2,B)} \; .
\EN  

In terms of our notations, the braking due to the solar wind is given by:
\EQ
\label{Gamma1}
\gamma = \gamma_\mathrm{\Sun} \times
\begin{cases}
\left(J_* B_* / x_2 \right)^{2/3} B^{4/3} & \mathrm{if} \quad x_2 > J_* \\
B^2 & \mathrm{else}
\end{cases}
\; ,
\EN
where $J_* = \alpha_0 \Omega_* / \Omega_\Sun$ and $\gamma_\Sun = 1 / t_{w\Sun}$. The magnetic field can then be taken to be linearly proportional to the angular rotation or to saturate above threshold. Results in this paper are obtained by integrating Eq. (\ref{System2}) using Heun's method \citep{Kloeden92}.

\section{Initial Gamma distribution}
\label{InitKeppens}
\begin{figure}[h]
\includegraphics[scale=0.4,clip]{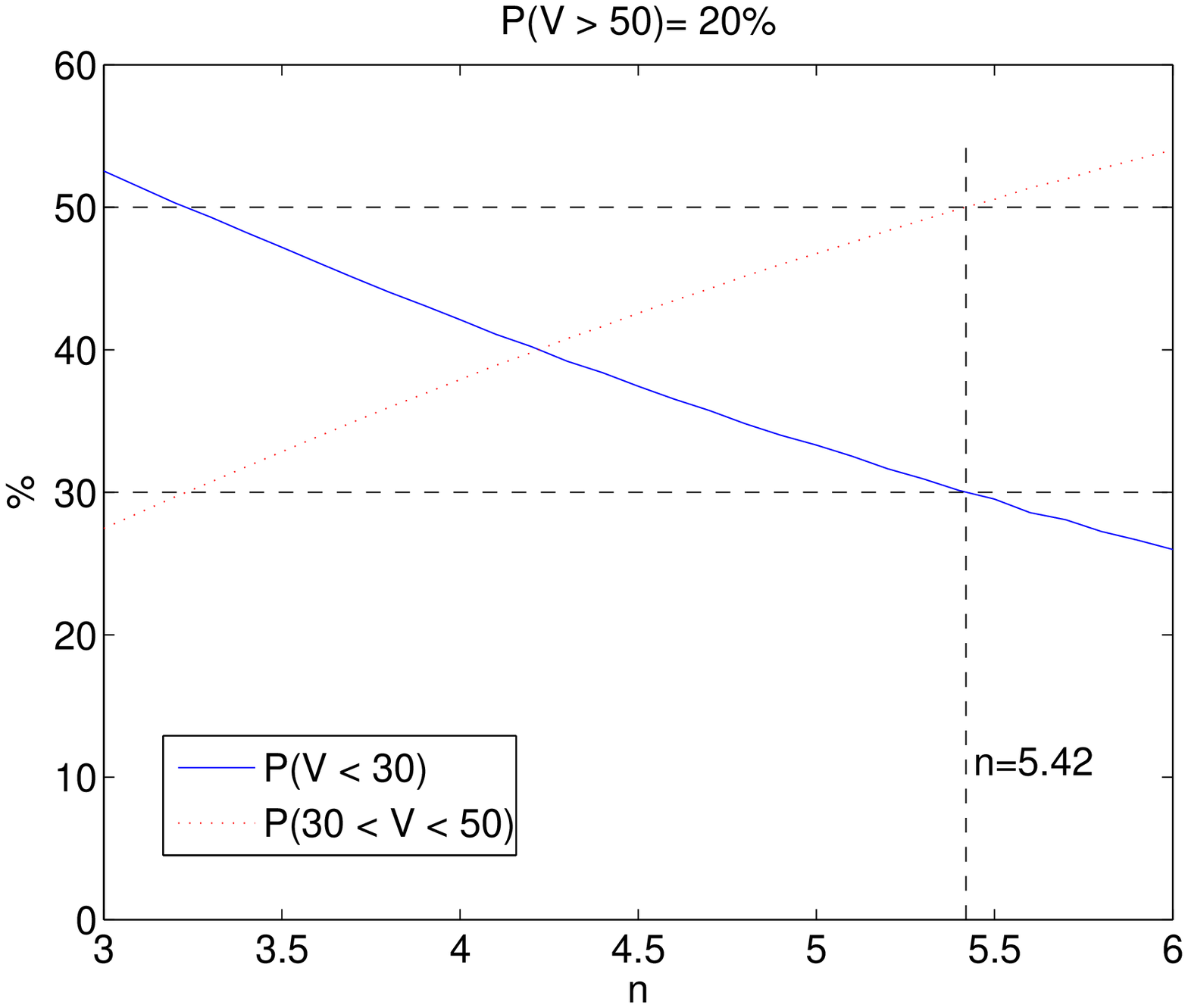}
\includegraphics[scale=0.4,clip]{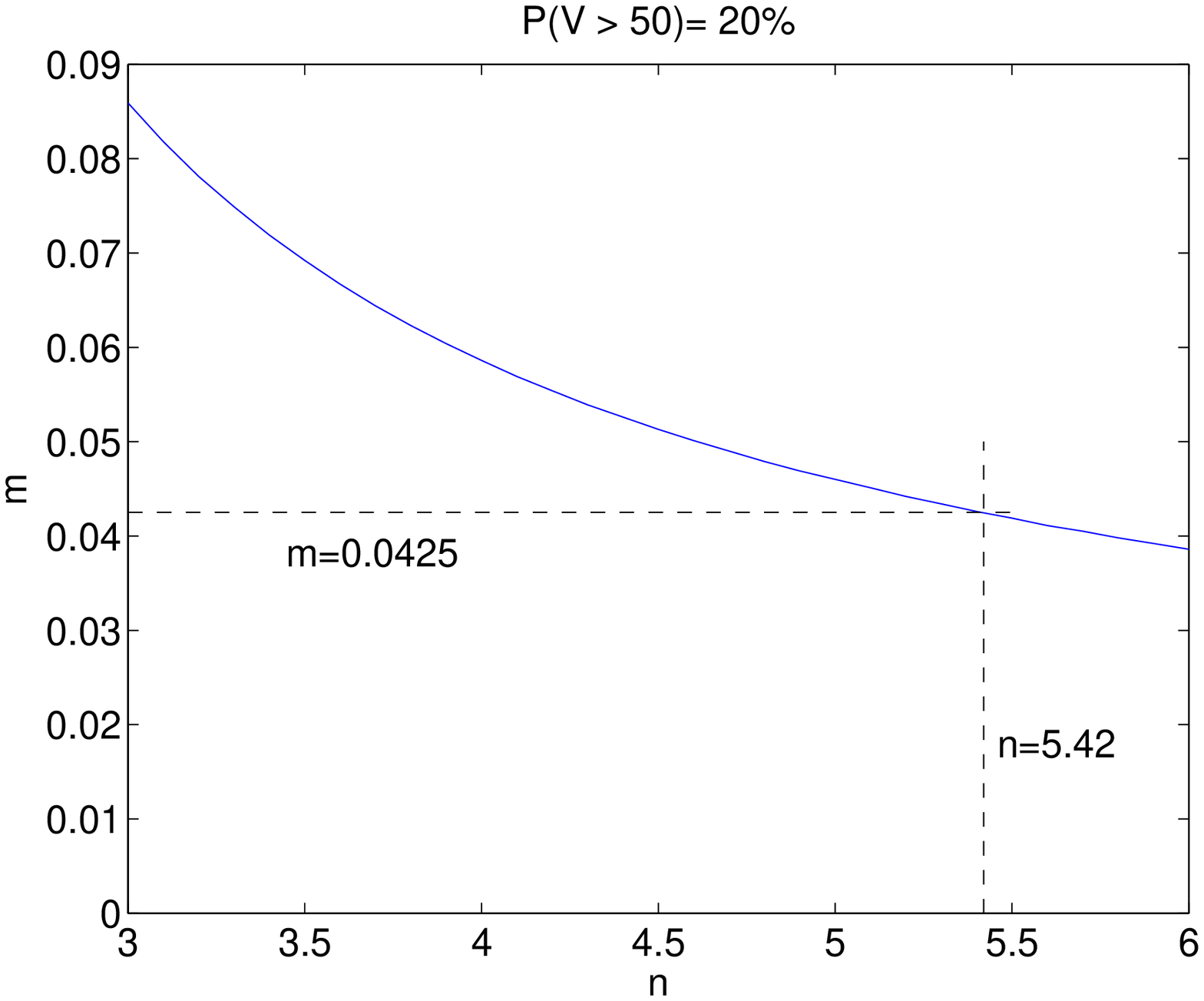}
\caption{\label{Pourc30} Choice of the values of $\alpha$ and $m$ when the percentage of slowly rotating stars is fixed to 30.}
\end{figure}

To compare our results with those of \citet{Keppens95}, we choose a family of initial distribution characterized by two parameters $n$ and $m$:
\EQ
P(x) = \frac{A}{x^n} \exp\left[\frac{-1}{(mx)^2}\right] \; .
\label{ExpStr}
\EN
The results of \citet{Keppens95} suggest that in the linear dynamo case (without saturation), the populations of $30 \, , \, 50 \; \mbox{and}  \; 20 \%$ of stars are observed for $V_{eq}$ lying in the intervals $0-30 \mbox{km}.\mbox{s}^{-1}$, $30-50 \mbox{km}.\mbox{s}^{-1}$ and $> 50 \mbox{km}.\mbox{s}^{-1}$, respectively. Figure \ref{Pourc30} shows the evolution of the percentages with the parameters $\alpha$ and $m$ when the percentage of stars with velocity under $30 \mbox{km}.\mbox{s}^{-1}$ is fixed at 30\%. The best fit to the initial distribution used by \citet{Keppens95} ($30 \, , \, 50 \; \mbox{and}  \; 20 \%$ for $V_{eq}$ in the intervals $0-30 \mbox{km}.\mbox{s}^{-1}$, $30-50 \mbox{km}.\mbox{s}^{-1}$ and $> 50 \mbox{km}.\mbox{s}^{-1}$, respectively) is obtained for $n =5.42$ and $m=0.0425$.

\begin{figure}[h]
\includegraphics[scale=0.55,clip]{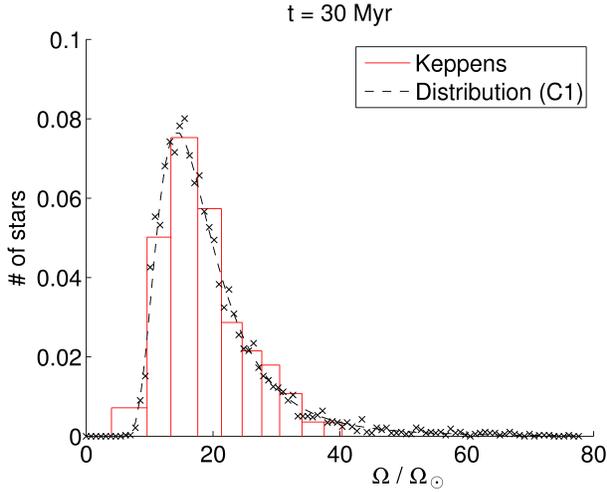}
\caption{\label{InitKeppens2} Comparison of the initial distribution calculated by \citet{Keppens95} and approximated by distribution (\ref{ExpStr}).}
\end{figure}

Figure \ref{InitKeppens2} compares the initial distribution we use at $30 \mbox{Myr}$ for simulations and the one computed by \citet{Keppens95}.

\section{Effect of the line of sight}
\label{AngleEffect}
Using data from \cite{Soderblom93}, we construct Table \ref{Soderblom} which shows the percentage of the population of stars having $v \sin i$ in a given velocity band.
\begin{table}[h] 
\begin{tabular}{|c|c|c|c|} 
\hline 
&   $t=50\, \mbox{Myr}$  & $t=70\, \mbox{Myr}$ &  $t=600\, \mbox{Myr}$ \\ 
\hline 
$v \sin i < 10$ & 23.4 \% & 51.4 \% & 86.2 \% \\
\hline 
$10 < v \sin i < 30$ & 38.3 \% & 37.5 \% & 13.8 \% \\
\hline 
$30 < v \sin i < 50$ & 6.4 \% & 7 \% & 0 \% \\
\hline 
$v \sin i > 50$ & 31.9 \% & 4.1 \% & 0 \% \\
\hline 
\end{tabular} 
\caption{\label{Soderblom} Percentages of the population of stars having a projected velocity $v \sin i$ inferred from the data in \cite{Soderblom93}.}
\end{table} 

The problem when using these data is that it overestimates the number of slow rotators (as $\sin i \leq 1$, it gives only a lower bound on the equatorial velocity of stars). To compensate this $\sin i$ factor, we assume that the distribution of the angle $i$ is uniformly distributed between $0$ and $\pi/2$. This means that the probability distribution of the angle is given by $P(i)=2/\pi$ for $0\leq i \leq \pi/2$. The probability of having an equatorial velocity between $v_1$ and $v_2$ is then computed as:
\EQA
\label{AngleCorrec}
P(v_1 \leq V_{eq} \leq v_2) &=& \int_{v_1 \leq x/\sin i \leq v_2} P(x=v\sin i) P(i) \, dx \, di \\ \nonumber
&=& \int_0^{v_2} dx P(x) \int_{\arcsin(x/v_2)}^{min(\arcsin(x/v_1),\pi/2)} di \, P(i) \\ \nonumber
&=& \frac{2}{\pi} \int_0^{v_2} dx P(x) \left[min(\arcsin(x/v_1),\pi/2)-\arcsin(x/v_2)\right] \; .
\ENA
Using Eq. (\ref{AngleCorrec}), we construct Table \ref{CorrecTable} which shows the probability of having different equatorial velocities.
\begin{table}[h] 
\begin{tabular}{|c|c|c|c|} 
\hline 
&   $t=50\, \mbox{Myr}$  & $t=70\, \mbox{Myr}$ &  $t=600\, \mbox{Myr}$ \\ 
\hline 
$V_{eq} < 10$ & 15.6 \% & 34.3 \% & 57.5 \% \\
\hline 
$10 < V_{eq} \sin i < 30$ & 29.6 \% & 35 \% & 28.7 \% \\
\hline 
$30 < V_{eq} < 50$ & 9.3 \% & 11.5 \% & 5.6 \% \\
\hline 
$V_{eq} > 50$ & 45.5 \% & 19.2 \% & 8.2 \% \\
\hline 
\end{tabular} 
\caption{\label{CorrecTable} Corrected table using formula (\ref{AngleCorrec}) for the percentage of stars having a given equatorial velocity.}
\end{table} 
We see that in this case, the number of slow rotators is decreased which agrees better with the numerical results of Table \ref{TableQs}, for instance for the $\sigma^2=0.1$ case. However these results are still not in perfect agreement with our theoretical predictions. First, one can see that the observations still show a larger proportion of fast rotators than that predicted in our stochastic model. Secondly, the observational data suggest that the rotation distribution is bimodal with a large proportion of rotators with $V_{eq} > 50$ and $10 < V_{eq} \sin i < 30$ and a gap in the range $30 < V_{eq} < 50$. This could be due to a lack of measurement in the range $30 < V_{eq} < 50$; more measurements would be needed to be included in order to assess the relevance of this observation. If this is not a measurement artifact, other physical ingredients are needed in order to explain these observations;  for instance, it has been proposed that the distribution at ZAMS should be bimodal due to a number of stars being prevented from spin-up during the PMS due to coupling to their accretion disks. 

An alternative approach is to assume a given value of the angle between the rotation rate and the line of sight and compute the projected velocities from the results of our numerical simulations. This is done in Table \ref{TableQs2} which has been constructed by using the equatorial velocity of Table \ref{TableQs} and projecting the velocity with an assumed angle $i = \pi/4$. The main difference between  Table \ref{TableQs} and Table \ref{TableQs2} is that the number of (observed) slow rotators has been severely increased in the latter case compared to the former one. For instance, for the case $\sigma^2 =0.1$, it changes from $2\%$ to $10.9\%$ at $50 \, \mbox{Myr}$. It should however be noted that this is still quite far from the observed percentage given in  Table \ref{Soderblom} which is $23.4\%$.
\begin{table}[h] 
\begin{tabular}{|c|c|c|c|c|c|} 
\hline 
&  & $t=30\, \mbox{Myr}$ &  $t=50\, \mbox{Myr}$  & $t=70\, \mbox{Myr}$ &  $t=600\, \mbox{Myr}$ \\ 
\hline 
\hline 
\multirow{4}*{Linear Case} & $V_{eq} \sin i < 10$ & 0.0 \%  & 1.0 \%  & 1.7 \%  & 77.5 \%   \\ \cline{2-6} 
 & $10 < V_{eq} \sin i < 30 $ & 67.5 \%  & 96.7 \%  & 96.4 \%  & 22.5 \%   \\ \cline{2-6} 
& $30 < V_{eq} \sin i < 50 $ & 26.4 \%  & 2.3 \%  & 1.9 \%  & 0.0 \%   \\ \cline{2-6} 
 & $V_{eq} \sin i > 50 $ & 6.1 \%  & 0.0 \%  & 0.0 \%  & 0.0 \%   \\  
\hline 
\hline 
\multirow{4}*{$Q_s = 20$} & $V_{eq} \sin i < 10$ & 0.0 \%  & 1.0 \%  & 1.7 \%  & 77.6 \%   \\ \cline{2-6} 
 & $10 < V_{eq} \sin i < 30 $ & 67.5 \%  & 95.4 \%  & 96.0 \%  & 22.4 \%   \\ \cline{2-6} 
& $30 < V_{eq} \sin i < 50 $ & 26.4 \%  & 2.3 \%  & 1.5 \%  & 0.0 \%   \\ \cline{2-6} 
 & $V_{eq} \sin i > 50 $ & 6.1 \%  & 1.3 \%  & 0.8 \%  & 0.0 \%   \\  
\hline 
\hline 
\multirow{4}*{$Q_s = 10$} & $V_{eq} \sin i < 10$ & 0.0 \%  & 1.4 \%  & 1.7 \%  & 73.5 \%   \\ \cline{2-6} 
 & $10 < V_{eq} \sin i < 30 $ & 67.5 \%  & 77.4 \%  & 79.6 \%  & 24.3 \%   \\ \cline{2-6} 
& $30 < V_{eq} \sin i < 50 $ & 26.4 \%  & 15.4 \%  & 13.2 \%  & 2.2 \%   \\ \cline{2-6} 
 & $V_{eq} \sin i > 50 $ & 6.1 \%  & 5.8 \%  & 5.5 \%  & 0.0 \%   \\  
\hline 
\hline 
\multirow{4}*{$Q_s = 5$} & $V_{eq} \sin i < 10$ & 0.0 \%  & 100.0 \%  & 100.0 \%  & 100.0 \%   \\ \cline{2-6} 
 & $10 < V_{eq} \sin i < 30 $ & 67.5 \%  & 0.0 \%  & 0.0 \%  & 0.0 \%   \\ \cline{2-6} 
& $30 < V_{eq} \sin i < 50 $ & 26.4 \%  & 0.0 \%  & 0.0 \%  & 0.0 \%   \\ \cline{2-6} 
 & $V_{eq} \sin i > 50 $ & 6.1 \%  & 0.0 \%  & 0.0 \%  & 0.0 \%   \\  
\hline 
\hline 
\multirow{4}*{$\sigma^2 =5$} & $V_{eq} \sin i  < 10$ & 0.0 \%  & 6.2 \%  & 8.5 \%  & 70.6 \%   \\ \cline{2-6} 
 & $10 < V_{eq} \sin i  < 30 $ & 67.5 \%  & 87.9 \%  & 86.7 \%  & 29.3 \%   \\ \cline{2-6} 
& $30 < V_{eq} \sin i < 50 $ & 26.4 \%  & 5.4 \%  & 4.5 \%  & 0.1 \%   \\ \cline{2-6} 
 & $V_{eq} \sin i > 50 $ & 6.1 \%  & 0.5 \%  & 0.4 \%  & 0.0 \%   \\  
\hline 
\hline 
\multirow{4}*{$\sigma^2 =1$} & $V_{eq} \sin i < 10$ & 0.0 \%  & 9.7 \%  & 10.6 \%  & 54.8 \%   \\ \cline{2-6} 
 & $10 < V_{eq} \sin i < 30 $ & 67.5 \%  & 77.3 \%  & 77.5 \%  & 44.6 \%   \\ \cline{2-6} 
& $30 < V_{eq} \sin i < 50 $ & 26.4 \%  & 11.0 \%  & 10.2 \%  & 0.6 \%   \\ \cline{2-6} 
 & $V_{eq} \sin i > 50 $ & 6.1 \%  & 1.9 \%  & 1.7 \%  & 0.0 \%   \\  
\hline 
\hline 
\multirow{4}*{$\sigma^2 =0.1$} & $V_{eq} \sin i < 10$ & 0.0 \%  & 10.9 \%  & 11.7 \%  & 47.3 \%   \\ \cline{2-6} 
 & $10 < V_{eq} \sin i < 30 $ & 67.5 \%  & 72.5 \%  & 71.4 \%  & 49.3 \%   \\ \cline{2-6} 
& $30 < V_{eq} \sin i < 50 $ & 26.4 \%  & 13.9 \%  & 13.9 \%  & 2.9 \%   \\ \cline{2-6} 
 & $V_{eq} \sin i > 50 $ & 6.1 \%  & 2.7 \%  & 2.9 \%  & 0.5 \%   \\  
\hline 
\end{tabular} 
\caption{\label{TableQs2} Percentages of stars having a certain projected angular velocity (with an angle $i=\pi/4$) for different values of the saturation rate $Q_s$ and the noise parameter $\sigma$. The other parameters have been fixed to $\tau_w=300 \mathrm{Myr}$ and $\tau_c=20 \mathrm{Myr}$.}
\end{table} 

\end{appendix}

\bibliographystyle{apj}
\bibliography{../../Biblio/Bib_sun,../../Biblio/Bib_maths,../../Biblio/Bib_dynamo,../../Biblio/Bib_procstoc,../../Biblio/Bib_shear,Bib_SpinDown}

\begin{thebibliography}{12}
\expandafter\ifx\csname natexlab\endcsname\relax\def\natexlab#1{#1}\fi

\bibitem[{Barnes \& Sofia(1996)}]{Barnes96}
Barnes, S. \& Sofia, S. 1996, Astrophys. J., 462, 746

\bibitem[{Bavassano {et~al.}(2005)Bavassano, Bruno, \& D'Amicis}]{Bavassano05}
Bavassano, B., Bruno, R., \& D'Amicis, R. 2005, Annales Geophysicae, 23, 1025

\bibitem[{Belcher \& MacGregor(1976)}]{Belcher76}
Belcher, J.~W. \& MacGregor, K.~B. 1976, Astrophys. J, 210, 498

\bibitem[{Buzasi(1997)}]{Buzasi97}
Buzasi, D.~L. 1997, Astrophys. J., 484, 855

\bibitem[{Forg\'acs-Dajka \& Borkovits(2007)}]{Forgacs07}
Forg\'acs-Dajka, E. \& Borkovits, T. 2007, Mon. Not. R. Astron. Soc., 374, 282

\bibitem[{Kepens {et~al.}(1995)Kepens, MacGregor, \& Charbonneau}]{Keppens95}
Kepens, R., MacGregor, K.~B., \& Charbonneau, P. 1995, Astron. Astrophys., 294,
  469

\bibitem[{Kloeden \& Platen(1992)}]{Kloeden92}
Kloeden, P.~E. \& Platen, E. 1992, Numerical solution of stochastic
  differential equations (Springer-Verlag)

\bibitem[{MacGregor \& Brenner(1991)}]{MacGregor91}
MacGregor, K.~B. \& Brenner, M. 1991, Astrophys. J., 376, 204

\bibitem[{O'Dell {et~al.}(1995)O'Dell, Panagi, Hendry, \& {Collier
  Cameron}}]{ODell95}
O'Dell, M.~A., Panagi, P., Hendry, M.~A., \& {Collier Cameron}, A. 1995,
  Astron. Astrophys., 294, 715

\bibitem[{Skumanich(1972)}]{Skumanich72}
Skumanich, A. 1972, Astrophys. J, 171, 565

\bibitem[{Soderblom {et~al.}(1993)Soderblom, Stauffer, MacGregor, \&
  Jones}]{Soderblom93}
Soderblom, D.~R., Stauffer, J.~R., MacGregor, K.~B., \& Jones, B.~F. 1993,
  Astrophys. J, 409, 624

\bibitem[{Weber \& Davis(1967)}]{Weber67}
Weber, E.~J. \& Davis, L.~J. 1967, Astrophys. J, 148, 217

\end{thebibliography}

\end{document}